\documentclass[aps,pra,showpacs,superscriptaddress,groupedaddress,twocolumn,amssymb]{revtex4}
\usepackage[utf8]{inputenc}
\usepackage{graphicx}
\usepackage{dcolumn}
\usepackage{bm}
\usepackage{amsmath}
\usepackage{indentfirst}
\usepackage{float}
\usepackage[colorlinks]{hyperref}
\usepackage[dvipsnames]{xcolor}
\usepackage{braket}
\usepackage{amsthm}
\usepackage[none]{hyphenat}
\begin{document}

\title{Quantum Coherence, Holevo Bound and Quantum Discord}

\author{Ashutosh K. Goswami}

\email[]{ashutoshgoswami841@gmail.com}

\affiliation{Indian Institute of Science Education and Research Kolkata, Mohanpur 741 246, West Bengal, India}
\author{Prasanta K. Panigrahi}
\email[]{pprasanta@iiserkol.ac.in}
\affiliation{Indian Institute of Science Education and Research Kolkata, Mohanpur 741 246, West Bengal, India}

\date{\today}
\begin{abstract}
A physically motivated resource theory of coherence under operations which do not use coherence has been recently proposed. Any quantum process involving operations that do not use coherence can be simulated easily with a classical probability distribution. We consider the task of classical communication over a quantum channel and examine the loss in the classical capacity of the channel if the receiver is only allowed to perform operations that do not use coherence. We show that the loss in the classical capacity of the channel is given by the loss of coherence due to mixing according to the relative entropy measure. Subsequently, we show for bipartite separable quantum states, $\rho=\sum_x p_x \rho_{xa} \otimes \rho_{xb},$ basis dependent discord $\delta(A \leftarrow B)$ is bounded above by the loss of coherence $\sum_x p_x C_r(\rho_{xb}) -C_r (\sum_x p_x \rho_{xb})$ on Bob side. Using this relation, we derive the complementarity relation of quantum discord $D(A\leftarrow B)$ and accessible information $H(X:Y_{max})$ on Bob's side. We proved that the sum of quantum discord and accessible information is bounded by Holevo Bound.

\end{abstract}

\flushbottom
\maketitle
%
%
\thispagestyle{empty}


\section{Introduction}
Coherence is one of the basic features of quantum mechanics that distinguishes it from classical physics. Preserving and monitoring it are the fundamental challenges for the physical applications of quantum principles. Although it is fundamental to optics \cite{Wolf, Wolf1}, a rigorous mathematical formulation of coherence as a quantum resource has been achieved only recently \cite{Baumgratz, Aberg06}. 
  In a remarkable paper, Baumgratz, Cramer and Plenio \cite{Baumgratz, Hu1} defined incoherent states as the states whose density matrix written in a chosen basis $\big\{\Ket{i}\big\}$, do not have any off diagonal elements: $\rho=\sum_i p_i \Ket{i}\Bra{i}.$ Akin to entanglement not increasing under \emph{local operation and classical communication (LOCC)} \cite{Horodecki}, incoherent operations have been defined as operations that map an incoherent state to an incoherent state. Two different coherence measures, relative entropy and $l_1$ norm have been proposed, satisfying several monotonicity properties under incoherent operations \cite{Baumgratz}. A scheme has been demonstrated for evaluating skew information based coherence measure experimentally for finite dimensional system \cite{Girolami}. Operational meaning of coherence in terms of the tasks like coherence distillation and coherence cost under incoherent operation \cite{Winter} and noise needed to completely erase the coherence of a quantum state \cite{Uttam} has been achieved. Operational meaning of $l_1$ norm measure has also been explored extensively \cite{swapan}.
 It has been shown that coherence is a necessary resource for generating entanglement using incoherent operations \cite{Bera}. Distribution of coherence for multi-partite quantum systems and monogamy relations has been derived \cite{Chandra}. A unified theory  of quantum correlations and coherence has been achieved \cite{Kok}. 
Quantum discord is a measure of quantum correlations between two parties A and B \cite{Vedral, Zurek}. It is invariant under local unitary transformations and computing it is NP complete \cite{Yichen}. A basis independent measure of coherence for multipartite system has been proposed \cite{Yao} and shown to be equal to the distance measure of quantum discord \cite{Modi}. Interestingly, the upper bound on quantum discord that can be generated by incoherent operations has also been derived \cite{ Ma1}.

A physically motivated resource theory of coherence utilizing operations that do not use coherence has been initiated recently  \cite{Yadeen}. These are the operations, that do not use off-diagonal elements of a density matrix with respect to a reference basis $\{\Ket{i}\}.$ There exists at least one Kraus representation $\{\hat{K_l}\}$ for such operations such that $tr(\hat{K_l}\rho \hat{K_l}^\dagger) = tr(\hat{K_l} \rho^d \hat{K_l}^\dagger)$ holds for all $l$ and any density matrix $\rho,$ with $\rho^d = \sum_i  \Ket{i}\Bra{i}\rho\Ket{i}\Bra{i}$ \cite{Yadeen}. Since the action of these operations is limited to only diagonal elements of a density matrix, using them in quantum protocols lead to certain disadvantage or loss. In this paper, we consider classical communication over a quantum channel. The process of communication can be described as: Alice picks a state from the ensemble $\big\{\rho_1, \rho_2, ......, \rho_n \big\}$ according to the probability distribution $\big\{p_1, p_2, ........, p_n\big\}$ and sends it to Bob. Bob's task is to perform a measurement $Y$ and infer the state Alice has prepared. The Holevo bound $S(\sum_x p_x \rho_x) - \sum_x p_x s(\rho_x)$ is the upper bound on the information gain about $X$ after performing a measurement $Y$, known as classical capacity of the quantum channel \cite{Holevo}. This bound can be achieved in the asymptotic limit \cite{Holevo1, Benjamin}. It has been proved that if Bob is allowed to only perform operations, which do not use off diagonal elements of a density matrix, the loss in the classical capacity of the quantum channel is given by the decoherence or loss of coherence (due to mixing) according to the relative entropy measure of coherence. Loss of coherence due to mixing is the difference between coherence of formation of the ensemble $\{p_x, \rho_x\}$ and coherence of the mixed density matrix $\sum_x p_x \rho_x.$ 


Subsequently, we establish the relation between loss of coherence due to mixing and the basis dependent quantum discord \cite{Vedral, Zurek}. The basis dependent discord for a bipartite system $\rho_{AB}$ is the difference between the quantum mutual information $I(A : B)$ and classical mutual information $J(A:B)$ in a chosen basis $\{\Ket{i}\}$. The basis independent discord, the minimum value of the basis dependent discord over all bases, has been recognized as a measure of quantum correlation, which can be present even in separable density matrices. However, it does not qualify as a robust measure of quantum correlation as it can be increased under local operations \cite{Alexander}. This poses the question under which set of operations discord does not increases and a significant set of these operations has been recognized \cite{Yadeen}. We show  that for a separable system $\rho=\sum_x p_x \rho_{xa} \otimes \rho_{xb},$ basis dependent discord is upper bounded by the loss of coherence due to mixing on Bob's side; $\sum_x p_x C_r(\rho_{xb}) -C_r (\sum_x p_x \rho_{xb})$. We employed this relation to prove the complementarity of quantum discord and accessible information on Bob's side;
$$ D(A \leftarrow B) + H(X:Y_{max}) \leq \chi,$$
here, $Y_{max}$ is the optimal measurement for which accessible information is maximum. Thus, if Holevo bound is achievable by performing certain measurement on Bob's side, quantum discord will be equal to zero.


First, we briefly review the concept of quantum coherence, strictly incoherent operations and local quantum incoherent operation and classical communication (LQICC). 
\section{Coherence Measure}
A quantum state $\rho$ is said to be incoherent with respect to a basis $\{\Ket{i}\},$ if it can be written as,
$\rho = \sum_i p_i \Ket{i}\Bra{i}$ and a completely positive trace preserving map $\Delta$ is an incoherent operation iff, $\Delta[\rho]$ is an incoherent quantum state, for all $\rho \in I,$ here $I$ is the set of all incoherent states. Based on this definition of incoherent quantum state and incoherent operations, Baumgratz et al proposed following monotonicity properties, any valid coherence measure $C(\rho)$ must satisfy   \cite{Baumgratz},
\begin{itemize}
\item  $C(\rho) \geq 0$ and $C(\rho)=0$ iff $\rho \in I. $
\item  $ C(\rho)$ must not increase under an incoherent operation; $C(\Delta[\rho]) \leq C(\rho).$
\item  $ C(\rho)$ must not increase under selective measurements; $ \sum_i p_i C(\rho_i) \leq C(\rho),$ where $\rho_i = \hat{K_i} \rho \hat{K_i^\dagger}/p_i,$ $p_i = tr(\hat{K_i} \rho \hat{K_i^\dagger}),$ with $\hat{K_i}$ being an incoherent operation.
\item $C(\rho)$ must not increase under mixing of quantum states; $\sum_i p_i C(\rho_i) \leq C(\sum_i p_i \rho_i).$
\end{itemize}
Following are two measures of coherence satisfying all four constraints;\\
\begin{itemize}
\item \emph{Relative entropy Based Measure:} $C_r(\rho)= min_{\sigma \in I} S(\rho||\sigma) = S(\rho^d)-s(\rho),$ here, $S(\rho^d)$ is the diagonal part of density matrix $\rho$ with respect to the reference basis, $\{\Ket{i}\}.$\\
\item \emph{$l_1$ norm based measure:} $ C_{l_1}(\rho) = \sum_{i, j, i\neq j} |\rho_{ij}|$
\end{itemize}
\textbf{Incoherent and strictly incoherent operations:}
Any incoherent Kraus operator can be written as, $\hat{K_\mu} = \sum_i c(i) \Ket{f_\mu (i)} \Bra{i},$ here, $\Ket{f_\mu (i)}$ is again an element of the incoherent basis set $\{\Ket{i}\}.$ The operators $\hat{K_\mu}$ and $\hat{K_\mu}^\dagger$ are both incoherent if and only if mapping $f_\mu (i)$ is one to one. These operators are known as strictly incoherent operation $\hat{K_l}$. Strictly incoherent operations are identified as operations which do not use quantum coherence of a resource state i.e., for any density matrix $\rho,$ relation $tr(\hat{K_l}\rho \hat{K_l}^\dagger) = tr(\hat{K_l}\rho^d \hat{K_l}^\dagger)$ holds for any strictly incoherent operator $\hat{K_l}$ \cite{Winter, Yadeen}. Although an incoherent operation can never generate coherence, implementing incoherent operations that use coherence require quantum instruments (beam splitter) similar to the coherent generating operations \cite{Yadeen}. 

\textbf{A Local Coherence Measure:}
A notion of local coherence for a bipartite density matrix $\rho_{AB}$ has been introduced \cite{Chitamber}. A bipartite density matrix $\sigma_{AB}$ will have no local coherence or said to be \emph{locally incoherent} on Bob side with respect to a chosen basis $\{\Ket{i}\}$, if it can be written as,
\begin{equation*}
     \sigma_{AB}= \sum_k p_k \sigma_A^k \otimes \xi_B^k,\end{equation*} here, $\xi_B^k$ is an incoherent state: $\xi_B^k=\sum_i p_i \Ket{i}\Bra{i}.$
 Operations which map a locally incoherent state to a locally incoherent state are local quantum incoherent operation and classical communication $(LQICC)$, which allows any local operation on Alice's side, but only incoherent operations on Bob's side.  A measure of local coherence has been proposed satisfying above mentioned four properties under LQICC operations is given by,
 \begin{equation*}  C_r^{A|B}(\rho_{AB})= min_{\sigma_{AB} \in QI} S(\rho_{AB}||\sigma_{AB}),\end{equation*}
here, $QI$ is the set of all locally incoherent states.
It can also be written as,
\begin{equation*}
     C_r^{A|B}(\rho_{AB})= S(\rho_{AB}^{d_B})- S(\rho_{AB}),
\end{equation*}
with $\rho_{AB}^{d_B}$ being block diagonal part of density matrix $\rho_{AB}$ in the chosen basis $\{\Ket{i}\}$ on Bob's side.
\section{Quantum Coherence and Holevo Bound}
 We consider an ensemble of density matrices $\{\rho_x\}=\{\rho_1, \rho_1, \dots \rho_n\}$. Alice chooses a quantum state from this set, according to the probability distribution  $\{p_x\}=\{p_1, p_2, \dots p_n\}$ and sends it to Bob.  The density matrix of the system for Bob  reads, $ \rho = \sum_x p_x \rho_x.$ Bob's task is to perform a measurement $Y$ and deduce which state Alice prepared. The maximum information that Bob can gain about the state prepared by Alice is the classical capacity of a quantum channel.\\
 We, first prove the following lemma:\\
 
\textbf{Lemma 1:} If Bob is restricted to perform only strictly incoherent operations, the maximum information that Bob gains about $X$ is given by, $$H(X:Y)= S(\rho^d)-\sum_x p_x S(\rho_x^d).$$

\textbf{Proof:}
The post-measurement state when Bob performs a projective measurement in the reference basis $Y=\{\Ket{y}\},$ on the density matrix $\rho = \sum_x p_x \rho_x$ reads, $\rho^{d}= \sum_i  \Bra{y}\sum_x p_x\rho_x\Ket{y} \Ket{y}\Bra{y},$ here $\rho^d$ is diagonal part of $\rho$ in the basis $\{\Ket{y}\}.$
The information that Bob gains after performing measurement $Y$ is, $H(Y)=S(\rho^{d})$ and the average information gain if Bob knows the value of the random variable X before performing measurement, $ H(Y|X)=\sum_x p_x H(Y|X=x)=\sum_x p_x S(\rho_x^d).$ Thus, the mutual information between the variables X and Y reads,
\begin{equation*}
     H(X:Y)= H(Y)-H(Y|X)= S(\rho^d)-\sum_x p_x S(\rho_x^d),\end{equation*}
Now, if Bob performs any measurement using  strictly incoherent Kruas operators,  $Y= \{ \hat{K_l}\},$  mutual information will never exceed $S(\rho^d)-\sum_x p_x S(\rho_x^d),$ since $p_l= tr(\hat{K_l} \rho \hat{K_l}^\dagger)= tr(\hat{K_l} \rho^d \hat{K_l}^\dagger),$ where $\rho^d$ is the state of the system after performing a measurement in basis $\{\Ket{y}\}.$ This shows probabilities $\{p_l\}$'s can be obtained from the post-processing of the state $\rho^d,$ hence must not contain more information than $S(\rho^d)-\sum_x p_x S(\rho_x^d)$ about $X.$ This completes the proof. \qed

\textbf{Theorem 1:} If receiver Bob is only allowed to perform strictly incoherent operations, decrease in classical capacity of a quantum channel is equal to the loss of coherence due to mixing.

\textbf{Proof:}
As is known, the accessible information to Bob about $X,$ after performing a POVM can never exceed Holevo's quantity, $\chi = S(\sum_x p_x \rho_x)-\sum_x p_x S(\rho_x);$  $\chi$ is also known as quantum mutual information \cite{Vedral1} and this Bound is achievable in asymptotic limit \cite{Benjamin, Holevo1}.


It is evident from \emph{lemma 1}, if $Y$ is a strictly incoherent operation, Holevo's quantity for this set of operations is given by, $S(\rho^d)-\sum_x p_x S(\rho_x^d).$ Thus, in the paradigm of strictly incoherent operation, there will be a permanent loss in the classical capacity of the quantum channel $(IL),$ which is given by,
\begin{equation} \begin{split} IL = \chi - H(X:Y) \implies S(\rho)-\sum_x p_x S(\rho_x)  \\ -S(\rho^d)+\sum_x p_x S(\rho_x^d), \end{split} \end{equation}
Interestingly, this is equivalent to the loss of coherence or decoherence $(CL)$ due to mixing with respect to the basis $\{\Ket{y}\}$ according to the relative entropy measure of coherence,
 \begin{equation}
  CL= \sum_x p_x C_{r}(\rho_x) - C_{r}(\sum_x p_x \rho_x).\end{equation}
Hence, we have,

\begin{equation}
IL = CL.
\end{equation}
      

Thus, decoherence of quantum systems due to mixing is equal to the loss in classical capacity of a quantum channel when the observer has an instrument that can only perform strictly incoherent operations ($SI$).  
For a coherence erasing channel $\rho \rightarrow  \rho' = \sum p_i U_i \rho U_i^{\dagger};$ \cite{Uttam} with $\rho'$ is diagonal in the reference basis, the loss in the classical capacity of the channel under strictly incoherent operations is equal to $C_r(\rho).$  \qed

It is evident from \emph{theorem 1} that for a pure state decomposition $\{p_x, \Ket{\psi_x}\}$ of the density matrix $\rho,$ average coherence or coherence of formation for $\rho$ will be minimum, when the loss of information $IL$ after performing a measurement in the reference basis $\{\Ket{y}\}$ is minimum.


%

\section{Quantum Coherence and Discord}
 For a bipartite system $\rho_{AB}$, the quantum mutual information between A and B reads,
    $$ I(A : B)= S(\rho_A) + S(\rho_B) - S (\rho_{AB}), $$
   However, after performing a local projective measurement Bob can retrieve only $J(A : B)$ information about subsystem A,
  $$ J (A : B) = S(\rho_A) - S(\rho_{AB}|\big\{\Pi_i^B\big\}),$$
   Similar to lemma $1,$ it can be proved that $J(A:B)$ is the maximum achievable information about $A,$ by performing any measurement involving only strictly incoherent Kraus operators on Bob's side. Hence, basis dependent discord $\delta(A \leftarrow B) = I(A:B) - J(A:B)$ is the loss of the mutual information under strictly incoherent operation on Bob's system.
 Discord, a measure of quantumness of correlations is the minimum value of $\delta(A \leftarrow B)$ over all projectors,\\
  $$D(A\leftarrow B) = min_{\{\Ket{i}\}}(I(A:B) - J(A:B))$$
  
Basis dependent discord, $\delta(A \leftarrow B)$ can also be written as \cite{Ma1, Yadeen}, 
  \begin{equation}
  \delta(A \leftarrow B) = C_r^{(A|B)}(\rho_{AB}) - C_{rel.ent}(\rho_B),
  \end{equation}
Thus, the difference between quantum mutual information and classical mutual information is equal to the difference between local coherence on Bob's side given by the LQICC monotone $C_r^{A|B}$ and the relative entropy measure of coherence of the reduced density matrix $\rho_B$. The local coherence on Bob's side given by LQICC monotone is the coherence of Bob's system in a multi-partite setting, while $C_{r}(\rho_B)$ is the coherence of subsystem $B,$ when sub-system $A$ has been discarded. For a separable bipartite density matrix shared between Alice and Bob $\rho_{AB}=\sum_x p_x \rho_{xa} \otimes \rho_{xb}$ \cite{Peres}, it is obvious that discarding Alice's system leads to the mixing of the quantum states $ \{\rho_{xb}\};$ $\rho_{B}=\sum_x p_x \rho_{xb}$ on Bob's side. Thus, we expect the $C_r^{(A|B)}$ will be bounded by average coherence on Bob's system; $C_r^{(A|B)}(\rho_{AB})\leq \sum_x p_x C_r(\rho_{xb})$. This has been proved in the following lemma:

\textbf{Lemma 2:} For the quantum state, $\rho_{AB}=\sum_x p_x (\rho_{xa}\otimes \rho_{xb})$ following inequality holds, $$ C_r^{A|B}(\rho_{AB}) \leq C_{avgB}=\sum_x p_x C_r(\rho_{xb}),$$ here both $C_r^{A|B}$ and  $C_r,$ measured with respect to the same basis $\{\Ket{i}_b\}$ on Bob's side and equality holds if and only if $\{\rho_{xa}\}'$s  are orthogonal to each other.

\textbf{Proof:} From convexity property of LQICC monotone,\\
$$C_r^{A|B}(\sum_x p_x (\rho_{xa}\otimes \rho_{xb})) \leq \sum_x p_x C_r^{A|B}(\rho_{xa}\otimes \rho_{xb})$$ From the definition, $C_r^{A|B}(\rho_{xa}\otimes \rho_{xb}) = S(\rho_{xa}\otimes \rho_{xb}^d) - S(\rho_{xa}\otimes \rho_{xb})=S(\rho_{xb}^d)-S(\rho_{xb}),$

\begin{equation*} \begin{split}
 C_r^{A|B}(\sum_x p_x (\rho_{xa}\otimes \rho_{xb})) \leq  \sum_i p_i ( S(\rho_{xb}^d)-S(\rho_{xb})) \\
 =  \sum_x p_x C_r(\rho_{xb}). \end{split} \end{equation*}
 equality holds if and only if, $\{\rho_{xa}\}$'s are orthogonal to each other. \qed

Using \emph{lemma 2}, we have following inequality for basis dependent quantum discord and loss of coherence ($CL$) due to mixing;
\begin{equation}
\delta(A \leftarrow B) \leq  \sum_x p_x C_r(\rho_{xb}) -C_r (\sum_x p_x \rho_{xb}),\end{equation}
Thus, basis independent discord is bounded above by the loss of coherence on Bob's side.
From theorem $1$ and equation $5,$ we have,
\begin{equation}\delta(A \leftarrow B) \leq IL_b \end{equation} 
As defined earlier, $IL_b= \chi - H(X:Y)$ is the loss of information after performing a projective measurement with respect to the reference basis $\{\Ket{y}\}$ on Bob's system. We have the following relation for the basis $Y_{max}$ for which the accessible information H(X:Y) is maximum,
\begin{equation}
\delta(A \leftarrow B)_{Y_{max}} \leq IL_{b_{min}}
\end{equation}
with $IL_{b_{min}}= \chi - H(X:Y_{max}).$ It's is easy to see that quantum discord, which is minimum of $\delta(A : B)$ over all possible measurements on Bob's side is also bounded by $IL_{b_{min}}.$ Thus, we have the following relation,
\begin{equation}
D(A \leftarrow B) + H(X:Y_{max}) \leq \chi
\end{equation}
This shows the complementarity of quantum discord and accessible information.
 We emphasize that Holevo bound appears naturally in the bipartite case and it's obvious from equation $8$ that if Holevo bound is achievable by performing a measurement on Bob's system, there will be no quantum discord; $D(A \leftarrow B)=0.$ We notice that equality for the loss of information and quantum discord has been established for classical-quantum states $\sum_x p_x (\Ket{x}_{a}\Bra{x}\otimes \rho_{xb})$ by Yao et al \cite{Yao13}, however our result  is more general and explicit.

 To conclude, firstly we considered the task of classical communication using quantum states $\{ p_x, \rho_x\}$ when the receiver is only allowed to perform strictly incoherent operation and demonstrated the equivalence between the loss of information about $x$  under strictly incoherence operations and the loss of coherence due to mixing. Thus, providing a information theoretic interpretation to the loss of coherence due to mixing; $CL = \sum_x p_x C_r(\rho_{x}) -C_r (\sum_x p_x \rho_{x}).$ 
 For a separable density matrix $\rho_{AB}= \sum_x p_ x \rho_{xa} \otimes \rho_{xb},$ it has been proved that basis dependent discord $\delta(A \leftarrow B)$ is bounded above by the loss of coherence $CL_{b}$ on Bob side. We utilized this inequality to show that the asymmetric quantum discord $D(A  \leftarrow B)$ is upper bounded by the minimum loss of information, $IL_{b_{min}}$ on Bob's system by performing the optimal quantum measurement $Y_{max}$. 
 This shows the complementarity of quantum discord, $D(A  \leftarrow B)$ and accessible information, $H(X:Y_{max})$ on Bob's ensemble. A direct implication of this result is if Holevo Bound is achievable by performing certain measurement on Bob's ensemble, quantum discord will be equal to zero.

 
\section{Acknowledgement}
    We acknowledge discussions with Subhayan Roy Moulick. 

\end{document}